\documentclass[pre,twocolumn]{revtex4}
\usepackage[english]{babel}
\usepackage{amsmath}
\usepackage{amssymb}
\usepackage{epsfig}
\usepackage{url}

\begin{document}
\title{Collisions of Light Bullets with Different Circular Polarizations}
\author{Victor P. Ruban}
\email{ruban@itp.ac.ru}
\affiliation{Landau Institute for Theoretical Physics RAS,
Chernogolovka, Moscow region, 142432 Russia}

\date{\today}

\begin{abstract}
Collisions of left- and right-polarized spatiotemporal
optical solitons have been numerically simulated for a
locally isotropic focusing Kerr medium with anomalous 
chromatic dispersion. The stable propagation of such
``light bullets'' in a moderate nonlinear regime is
ensured by a transverse parabolic profile of the 
refraction index in a multimode waveguide. The transverse
motion of centers of mass of wave packets in such systems
occurs on classical trajectories of a harmonic oscillator,
whereas the motion in the longitudinal direction is uniform. 
Therefore, collisions of two solitons can be not only head-on
but also tangential. An inelastic collision of two solitons
with opposite circular polarizations can result either in two
binary light bullets combining the left and right polarization
or in more complex bound systems.

\vspace{1mm}

\noindent DOI: 10.1134/S0021364024600691
\end{abstract}

\maketitle

\subsection*{Introduction}

In nonlinear optics, spatiotemporal optical solitons
have been studied for several decades (see [1--9] and
numerous references therein). Such stably propagating
coherent structures confined in three dimensions are
called light bullets [1--3]. Difficulty of this problem is
that the nonlinear Schrödinger equation used as a
basic mathematical model for a quasi-monochromatic light
wave in a Kerr medium does not have stable soliton
solutions in homogeneous three-dimensional space. 
A wave packet either spreads due to dispersion and 
diffraction or collapses due to nonlinearity (see [8--11]
and references therein). For this reason, an additional
stabilizing factor -- saturated nonlinearity, spatial
inhomogeneity, etc. -- is necessary (see, e.g., [4,6,7,12--22]). 
In particular, the effective transverse potential 
keeps a moderately nonlinear wave packet from
diffractive (transverse) spread in a multimode waveguide 
with a smooth (approximately parabolic) profile of the 
refractive index, and anomalous chromatic dispersion
in the longitudinal (temporal) direction is stably 
balanced by nonlinearity becoming effectively
one-dimensional [2,3,5,8,9]. In this case, the transverse 
degrees of freedom are not frozen, in contrast to the 
situation in single-mode optical fibers. The established 
balance results in the stability of light bullets
with not too high total energy. Furthermore, the center
of mass of such a soliton can be displaced in the transverse 
direction in a well-known way (on classical oscillator 
trajectories) so that its motion regime is intermediate 
between the rigid transverse fixation in a single-mode 
optical fiber and complete freedom in a homogeneous space.

However, the conventional scalar nonlinear
Schr\"od\-inger equation describes only one polarization
of light, linear or circular. A locally isotropic Kerr
medium allows the nonlinear interaction between
waves with two polarizations. In this case, the system
of two incoherently coupled nonlinear Schr\"odinger
equations for slow complex amplitudes of left and right
circularly polarized waves is appropriate [23]. Correspondingly, 
it is necessary to consider combined solitons, 
i.e., stable nonlinear solitary structures with both
circular polarizations generally in an arbitrary proportion. 
It is also important to know processes accompanying 
collisions of three-dimensional optical solitons
with different polarizations. To the best of my knowledge,
such problems were studied only for one-dimensional 
systems. Very nontrivial results were obtained for 
inelastic collisions of one-dimensional vector solitons 
(see, e.g., [24--26] and references therein). Even more
interesting dynamics should be expected in three dimensions.

A variational analysis performed in this work indicates 
the possibility of existence of moderately nonlinear 
binary light bullets. The three-dimensional numerical 
simulation confirms their stable nature. A number of
numerical experiments on head-on and tangential collisions
of three-dimensional light bullets with opposite circular
polarizations are also carried out. Processes accompanying
these inelastic collisions significantly differ from those
occurring in the case of the interaction of identically 
polarized solitons. It is found that, depending on the 
parameters, the collision can result in two binary light 
bullets propagating at changed velocities away from each other
and sometimes in the ``merging'' of colliding solitons,
forming a more complex nonstationary structure. Wave collapse
occurs at some initial parameters.

\subsection*{Main properties of the model}
    
We consider a transparent optical medium with the focusing 
Kerr nonlinearity and the dispersion relation of linear waves 
$k(\omega)=\sqrt{\varepsilon(\omega)}\omega/c$
with the anomalous dispersion $k''(\omega)<0$  
in a certain frequency range.
In particular, these properties are inherent in many
glasses. The known equation for the vector envelope of
a weakly nonlinear quasimonochromatic light wave
with the carrier frequency $\omega_0$ in the paraxial
approximation is taken as the main model. In the dimensional
variables, this equation has the form
\begin{eqnarray}
&&2k_0[-i\partial_\zeta-ik_0'\partial_t+k_0''\partial_t^2/2]{\bf E}
-\Delta_\perp {\bf E}\nonumber\\
&&\approx\frac{k_0^2}{\varepsilon(\omega_0)}
\Big[\tilde\varepsilon(x,y,\zeta){\bf E}
+\alpha|{\bf E}|^2{\bf E}+
\beta({\bf E}\cdot{\bf E}) {\bf E}^*\Big],
\end{eqnarray}   
where $\zeta$ is the coordinate along the beam, $k'_0=1/v_{\rm gr}$ is
the inverse group velocity of light in the medium, $k''_0$ is
the negative chromatic dispersion coefficient,
$\tilde\varepsilon(x,y,\zeta)$ is the small inhomogeneity of the relative
permittivity at the carrier frequency, and $\alpha(\omega_0)$ and 
$\beta(\omega_0)$ are the positive coefficients of nonlinearity. 
Let  $\tau=t-\zeta/v_{\rm gr}$ be the ``retarded'' time.
In terms of slow amplitudes $\psi_{1,2}(x,y,\tau,\zeta)$ 
of the left and right circular polarizations, 
the amplitude of the electric field is   
\begin{equation}
{\bf E}\approx \big[({\bf e}_x+i{\bf e}_y) \psi_1 
+ ({\bf e}_x-i{\bf e}_y) \psi_2 \big]/\sqrt{2}, 
\end{equation}    
and light is then described by two coupled scalar nonlinear 
Schr\"odinger equations [23], similar to a binary
Bose--Einstein condensate of cold atoms (with the
change of variables $\zeta \rightarrow t$, $\tau \rightarrow z$). 
The appropriate rescaling leads to the dimensionless system
\begin{equation}
i\frac{\partial \psi_{1,2}}{\partial \zeta}=
\Big[-\frac{1}{2}\Delta +U(x,y,\zeta)
-|\psi_{1,2}|^2 - g|\psi_{2,1}|^2\Big]\psi_{1,2},
\label{psi_12_eqs}
\end{equation}
where $\Delta=\partial_x^2+\partial_y^2+\partial_\tau^2$ 
is the three-dimensional Laplace operator in the ${\bf r}=(x,y,\tau)$ 
``coordinate'' space and $g=1+2\beta/\alpha$ 
is the cross-phase modulation parameter, which is about 2 in 
the typical case of fast nonlinear response. It is worth 
emphasizing that the nonlinear interaction between two components
is reduced to a simple noncoherent coupling through the
coefficient $g$ only in the basis in the form of circular
polarizations. Such a coupling conserves the amount
of each component $N_{1,2}=\int |\psi_{1,2}|^2 dx dy d\tau$ 
and makes it possible to apply the highly efficient numerical 
split-step Fourier method (see below). Nonlinear terms
corresponding to the so-called four-wave mixing
would remain in the system with two linear or two
elliptic polarizations. These terms are inconvenient
both analytically (because they result in transitions
between components) and numerically except for the
obvious reduction $\psi_2=\psi_1\exp(i\delta_0)$, 
which corresponds to the purely linear polarization.

The external parabolic potential $U\propto-\tilde\varepsilon(x,y,\zeta)$ 
can be represented in the compact matrix--vector form
\begin{equation}
U(x,y,\zeta)=\frac{{\bf r}\cdot \hat S(\zeta) {\bf r}}{2} 
-{\bf r}\cdot{\bf F}(\zeta) +U_0(\zeta),
\end{equation}
with the symmetric matrix $\hat S(\zeta)$, where only transverse
components are nonzero, the transverse vector ${\bf F}(\zeta)$,
and the scalar $U_0(\zeta)$. The functions ${\bf F}(\zeta)$ and 
$U_0(\zeta)$ include possible deviations of the central line 
of the waveguide from the axis $\zeta$ and variations of the depth
of the potential well.

The coupled nonlinear Schrödinger equations (3)
constitute a hydrodynamic system for two interacting
compressible perfect liquids whose flows are potential.
The transition to the corresponding hydrodynamic
variables -- densities $I_{1,2}$ and potentials of velocities 
$\varphi_{1,2}$ -- is ensured by the Madelung transformation
$$
\psi_s=\sqrt{I_s}\exp(i\varphi_s),\qquad s=1,2.
$$

A similar system of equations but with defocusing
nonlinearity was recently considered in [27]. In the
defocusing case, the positive ``hydrodynamic'' pressure
supports a delocalized density background, and
the main ``soft'' coherent structures are quantized vortices 
and domain walls between two circular polarizations. 
In this work, the case of focusing nonlinearity is
considered, and of main interest are soliton-like
objects where a negative hydrodynamic pressure (tension) 
is balanced by the ``quantum'' pressure (i.e., dispersion) 
in the longitudinal direction, whereas diffraction in the 
transverse direction slightly weakened by the hydrodynamic 
tension resists the transverse confining potential.

It is also important that the center of mass of any
localized distribution of liquids in the quadratic external
potential moves according to the equation of the classical 
oscillator (cf. [28]):
\begin{equation}
{\bf R}_{\rm c.m.}''=-\hat S(\zeta){\bf R}_{\rm c.m.}+{\bf F}(\zeta),
\end{equation}
where the double prime means the second derivative
with respect to the variable $\zeta$. In some cases, the trajectory 
of this motion can be quite nontrivial, e.g., in
the rotating anisotropic potential when the transition
to the rotating reference frame adds the Coriolis and
centrifugal forces. The periodic dependence of the
eigenvalues of the matrix $\hat S(\zeta)$ providing conditions for
the parametric resonance is also interesting. However,
even a conventional anisotropic two-dimensional
oscillator with the constant matrix 
$\hat S=\mbox{Diag}\{\kappa_1^2,\kappa_2^2,0\}$
``yields'' Lissajous figures. In this case, it is very substantial 
that the ``internal'' dynamics of the localized structure 
is ``insensitive'' to the motion of the center of mass. 
Indeed, let us consider a slightly ``truncated'' auxiliary 
system of equations without terms with ${\bf F}(\zeta)$
and $U_0(\zeta)$:
\begin{equation}
i\frac{\partial \Psi_{1,2}}{\partial \zeta}=\Big[-\frac{1}{2}\Delta +
\frac{{\bf r}\cdot \hat S(\zeta) {\bf r}}{2}
-|\Psi_{1,2}|^2 - g|\Psi_{2,1}|^2\Big]\Psi_{1,2}.
\label{Psi_12_eqs}
\end{equation}
It is easy to verify that the functions
\begin{equation}
\psi_s=\Psi_s({\bf r}-{\bf R}(\zeta), \zeta)
\exp[i ({\bf r}-{\bf R}(\zeta))\cdot{\bf P}(\zeta)-i\gamma(\zeta)]
\end{equation}
satisfy the complete system of equations (3) under the conditions
\begin{eqnarray}
{\bf R}'&=&{\bf P},\\
{\bf P}'&=&-\hat S(\zeta){\bf R}+{\bf F}(\zeta),\\
\gamma'&=&-\frac{{\bf P}^2}{2}+\frac{{\bf R}\cdot \hat S(\zeta) {\bf R}}{2}
-{\bf R}\cdot {\bf F}+U_0(\zeta).
\end{eqnarray}
The first two conditions are the equations of motion of
the classical oscillator in the presence of the driving
force ${\bf F}(\zeta)$. Without loss of generality, it can be
accepted that $\Psi_{1,2}$ correspond to the center of
mass at rest; then, it becomes obvious that the motion
of the center of mass does not affect the internal
dynamics, as in the case of the conventional nonlinear
Schr\"odinger equation [28].

\subsection*{Gaussian variational ansatz}

This internal dynamics itself can be fairly complex,
particularly, in the presence of both polarizations. In
particular, a combined soliton including two components
in the numbers $N_1$ and $N_2$ is certainly of interest.
Such binary light bullets have not yet been studied.
The properties of these objects in the simplest case of
two symmetric wave packets with coinciding centers
can be understood using the variational method. The
system of equations (6) corresponds to the Lagrangian
\begin{equation}
{\cal L}=\frac{i}{2}\int
\sum_{s=1,2}(\Psi_s^*\Psi'_s-\Psi_s\Psi_s'^*)dx dy d\tau
-{\cal H},
\end{equation}
where the Hamiltonian functional has the form
\begin{eqnarray}
{\cal H}&=&\frac{1}{2}\int\Big[\sum_{s=1,2} \big(|\nabla\Psi_s|^2
+{\bf r}\cdot \hat S  {\bf r}|\Psi_s|^2-|\Psi_s|^4\big)\nonumber \\
&&\qquad\qquad-2g|\Psi_1|^2 |\Psi_2|^2\Big]dx dy d\tau.
\end{eqnarray}
By analogy with the single-component nonlinear
Schr\"odinger equation, trial functions in the form of
Gaussian wave packets
\begin{eqnarray}
&&\Psi_s=\sqrt{N_s}\Big(\frac{\mbox{det}\hat A_s}{\pi^3}\Big)^{\frac{1}{4}}
\nonumber\\
&&\times\exp\Big(-\frac{{\bf r}\cdot \hat A_s(\zeta) {\bf r}}{2}
-i\frac{{\bf r}\cdot \hat B_s(\zeta) {\bf r}}{2} -i\phi_s(\zeta)\Big),
\end{eqnarray}
where $\hat A_s(\zeta)$ and $\hat B_s(\zeta)$ are unknown symmetric
matrices and $\phi_s(\zeta)$ are unknown phases, are substituted into
the Lagrangian. We emphasize that the matrices $\hat A_s(\zeta)$
and $\hat B_s(\zeta)$ are not generally assumed to be diagonal, in
contrast to most of the works where the variational
method is applied to the nonlinear Schr\"odinger equation.
Gaussian integrals are easily calculated in the
general form and lead to a Lagrangian system with a
finite number of degrees of freedom:
\begin{eqnarray}
4L&=&\sum_{s=1,2} \big[N_s\mbox{Tr}(\hat A_s^{-1}\hat B_s') +N_s\phi_s'\big] -4H,
\nonumber\\
4H&=&\sum_{s=1,2}\Big[N_s\mbox{Tr}(\hat A_s^{-1}\hat B_s^2+\hat A_s
+\hat A_s^{-1} \hat S)\nonumber\\
&&\qquad\quad
-\frac{2N_s^2}{\sqrt{8\pi^3}} \Big(\mbox{det}\hat A_s\Big)^{\frac{1}{2}}\Big]
\nonumber \\
&& -4 g\frac{N_1 N_2}{\sqrt{\pi^3}}
\big[\mbox{det}\big(\hat A_1^{-1}+\hat A_2^{-1}\big)\big]^{-\frac{1}{2}}.
\end{eqnarray}
The interaction between the polarizations is obviously
described by the last term.

Naturally, the variables $N_s$  are integrals of motion. 
The equations for the matrices $A_s(\zeta)$ and $\hat B_s(\zeta)$
have the structure
\begin{eqnarray}
&&\hat A_s'= (4/N_s)\hat A_s (\partial H/\partial \hat B_s)\hat A_s,\\
&&\hat B_s'=(4/N_s)(\partial H/\partial \hat A_s^{-1}).
\end{eqnarray}
The explicit form of partial derivatives with respect to
the elements of the matrix $\hat A_s^{-1}$  is easily determined
using the relations
$d\, \mbox{Tr}\hat M =-\mbox{Tr}(\hat M^2 d\hat M^{-1})$
and $d\, \mbox{det}\hat M=-\mbox{det}\hat M \mbox{Tr}(\hat M d\hat M^{-1})=
\mbox{det}\hat M \mbox{Tr}(\hat M^{-1} d\hat M)$,
where $\hat M$ is an arbitrary symmetric matrix. As a result,
the system of ordinary differential equations is
obtained in the compact matrix form
\begin{eqnarray}
&&\hat A_s'= \hat A_s\hat B_s+ \hat B_s\hat A_s,\\
&&\hat B_s'= \hat B_s^2 -\hat A_s^2 +\hat S+
\frac{N_s}{\sqrt{8\pi^3}}\Big(\mbox{det}\hat A_s\Big)^{\frac{1}{2}} \hat A_s
\nonumber\\
&& +2g\frac{N_{3-s}}{\sqrt{\pi^3}}
\big[\mbox{det}\big(\hat A_1^{-1}+\hat A_2^{-1}\big)\big]^{-\frac{1}{2}}
\big(\hat A_1^{-1}+\hat A_2^{-1}\big)^{-1}.
\end{eqnarray}
In the general form, this system is still too complex for
a detailed analysis. However, if only the diagonal
matrices $\hat S(\zeta)$, $\hat A_s(\zeta)$, and$\hat B_s(\zeta)$
are retained and $\hat A_s^{-1}=\mbox{Diag}\{a_s^2,b_s^2,c_s^2\}$
is set, the Newtonian dynamics for the ``vectors'' 
${\bf a}_s=(a_s,b_s,c_s)$ is obtained:
\begin{equation}
\tilde N_s {\bf a}_s''=-\partial \tilde W/\partial {\bf a}_s,
\label{Newton_abc}
\end{equation}
where slightly more convenient variables $\tilde N_s=N_s/\sqrt{8\pi^3}$
are introduced and the potential energy $\tilde W$ has the form
\begin{eqnarray}
&&\tilde W=\frac{2W}{\sqrt{8\pi^3}}\nonumber \\
&&=\sum_{s=1,2}\Big[\frac{\tilde N_s}{2}\Big(
\frac{1}{a_s^2}+\frac{1}{b_s^2}+\frac{1}{c_s^2}
+\kappa_1^2 a_s^2+\kappa_2^2 b_s^2\Big)
-\frac{\tilde N_s^2}{a_s b_s c_s}\Big]
\nonumber \\
&& \quad-
\frac{2 g\tilde N_1 \tilde N_2\sqrt{8}}
{\sqrt{(a_1^2+a_2^2)(b_1^2+b_2^2)(c_1^2+c_2^2)}}.
\end{eqnarray}
The corresponding dynamic system for the single-component 
nonlinear Schrödinger equation is studied in detail 
(see, e.g., [2,3,8,9,29]; it is noteworthy that
the longitudinal dependence in the diagonal variational
ansatz is sometimes taken in the form
$1/\cosh[\tau/c(\zeta)]$ rather than in the Gaussian form; the
potentials $W$  in this case are fairly similar).

If  $\kappa_1^2$ and $\kappa_2^2$ are independent of the evolution
variable $\zeta$, a local minimum of the function $\tilde W$ determines
the equilibrium configuration of a binary light bullet.
In the axisymmetric case (where $\kappa_1^2=\kappa_2^2=1$), it is
obvious that $a_s=b_s$, and only four of six variables
remain independent. In this case, the qualitative result
is the same as for the single-component soliton: the
function $\tilde W$ has a stable equilibrium position in a certain 
region of not too large parameters $\tilde N_1$ and $\tilde N_2$. If
$\tilde N_1>\tilde N_2$, then $a_1>a_2$ and $c_1>c_2$; i.e., the weaker
component is slightly more localized. This is natural because 
the nonlinear potential well $(-I_2-gI_1)$ for the weaker component
at $I_2\ll I_1$ is deeper than the well $(-I_1-gI_2)$ for 
the stronger component due to the parameter $g>1$.

\subsection*{Three-dimensional numerical simulation}

To verify the conclusions provided by the variational 
analysis and to study collisions of solitons, the
direct numerical simulation of the system of equations
(3) by the standard second-order split-step Fourier
method in the variable $\zeta$ was performed. The computational domain
had sizes $(4\pi)\times(4\pi)$ in the transverse $x$ and $y$ directions
and $8\pi$ or  $12\pi$ in the longitudinal $\tau$ direction. The lattice 
step $h=4\pi/128\sim 0.1$ in the $(x,y,\tau)$ ``space''
together with the evolution step $\delta\zeta=0.002$ ensured a
sufficient high resolution of smooth wave fields (in the
absence of collapse) and the conservation of the Hamiltonian 
with an accuracy of four to six decimal places
in the entire propagation distance of several hundreds
of dimensionless units of $\zeta$. The experiment was
stopped if a sharp and strong (several times) increase
in the maximum intensity began shortly before the
collapse although the spatial resolution in this case still
remained acceptable. For this reason, the process of
collapse was not simulated in this work.

\begin{figure}
\begin{center}
\epsfig{file=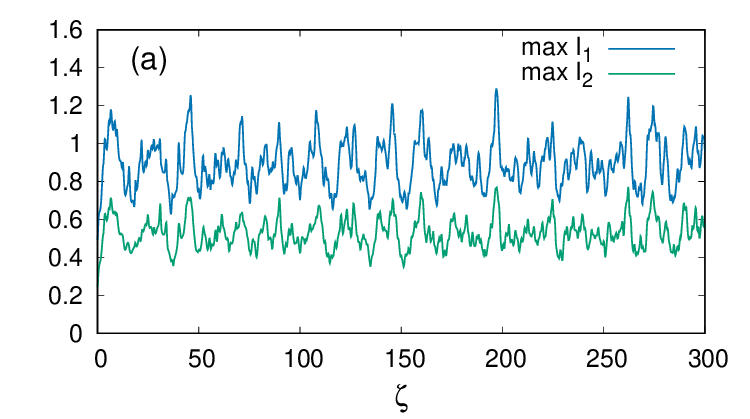, width=87mm}\\
\epsfig{file=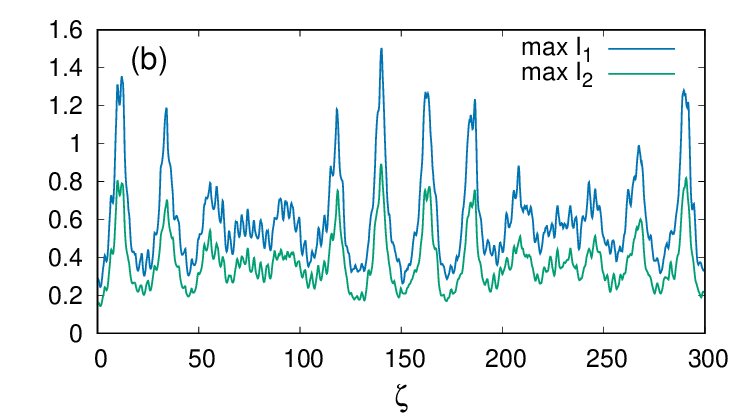, width=87mm}\\
\epsfig{file=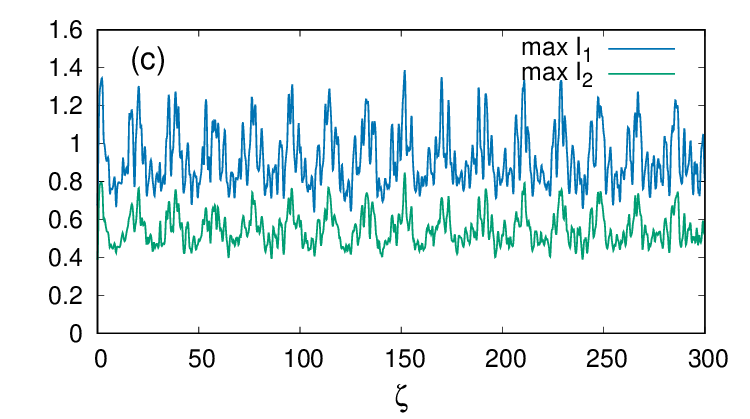, width=87mm}
\end{center}
\caption{Numerically obtained maximum intensities of the left and right
components of the nonlinear binary wave packet with $N_1=3.8$, $N_2=1.9$, 
and Gaussian initial conditions at $a_1^2=b_1^2=a_2^2=b_2^2=0.8$
and (a) $c_1^2=3.0$, $c_2^2=3.0$;
(b) $c_1^2=10.0$, $c_2^2=8.0$;
(c) $c_1^2=1.6$, $c_2^2=1.2$.
}
\label{max_I1_I2} 
\end{figure}

Figures 1a--1c show the numerically obtained
maximum intensities of the first and second components 
of the nonlinear binary wave packet with Gaussian 
initial conditions. As should be expected, detailed
coincidence with the Gaussian approximation was not
achieved. Degrees of freedom ``disregarded'' in the
Gaussian were rapidly excited in computer experiments 
due to the imperfection of the initial conditions.
However, it is important that these perturbations were
not enhanced at long distances; consequently, the
hypothesis of the stable propagation of binary light
bullets was confirmed by the numerical test.

At periodic dependences of the coefficients of the quadratic 
potential such as $\kappa_1^2(\zeta)=1+\epsilon_1\cos(2\zeta)$ 
and $\kappa_2^2(\zeta)=1+\epsilon_2\cos(2\zeta)$
with small parameters $\epsilon_1$ and $\epsilon_2$,
the parametric resonance occurs in the system, as well
as in the single-component case (see [30, 31] and
references therein). The computer simulation of Eqs. (3)
indeed demonstrated the parametric ``build-up'' in the
sizes of the binary soliton and a strong increase in its
energy up to the situation where the transverse size of
the soliton in the expansion phase exceeds the computational 
domain. However, since the performed numerical experiments
did not demonstrate any noticeable qualitative differences
between parametric resonances of solitons with one and
two polarizations, this subject was not considered in this work.

\begin{figure}
\begin{center}
\epsfig{file=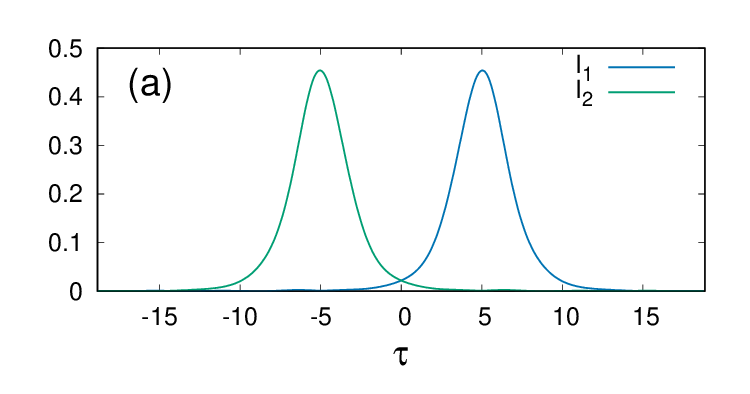, width=86mm}\\
\epsfig{file=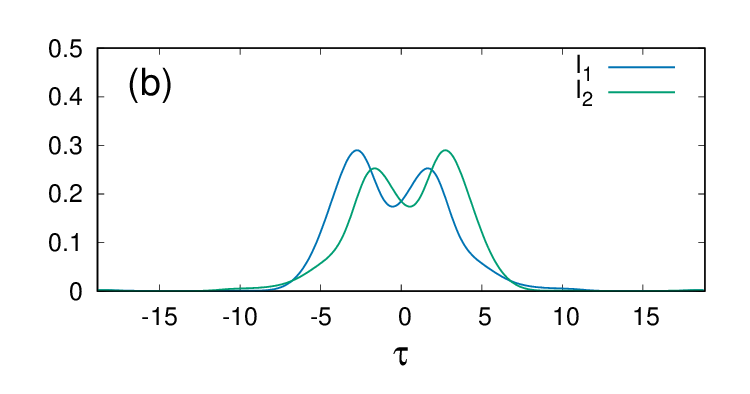, width=86mm}\\
\epsfig{file=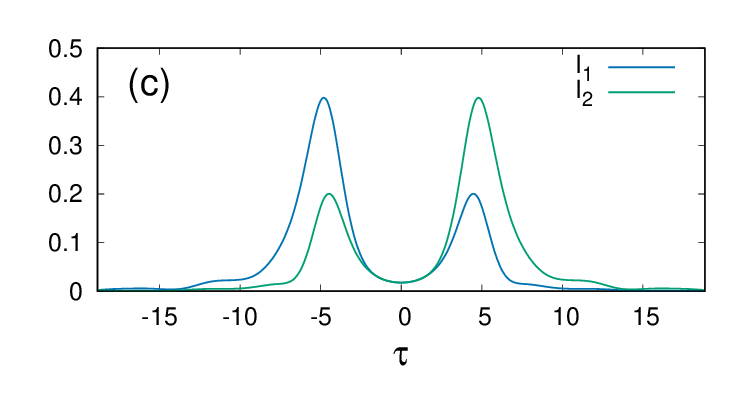, width=86mm}\\
\epsfig{file=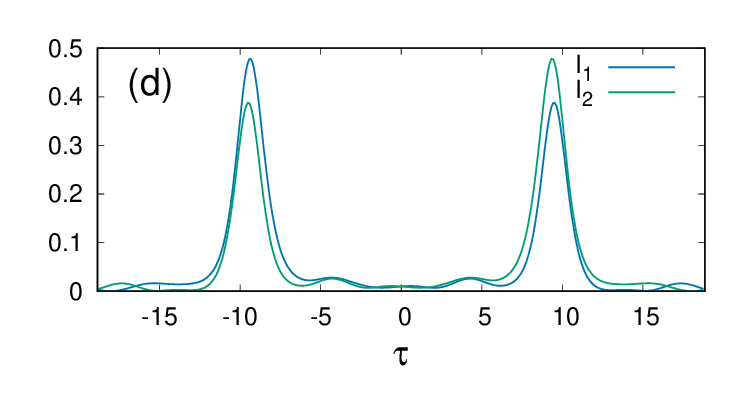, width=86mm}
\end{center}
\caption{Head-on collision ($R=0$) of light bullets at
their relative approaching velocity $v=0.1$. Both
intensities at $x=y=0$ are shown at propagation distances
$\zeta$ = (a) 20, (b) 60, (c) 80, and (d) 140.
}
\label{head-on-collision} 
\end{figure}

\begin{figure}
\begin{center}
\epsfig{file=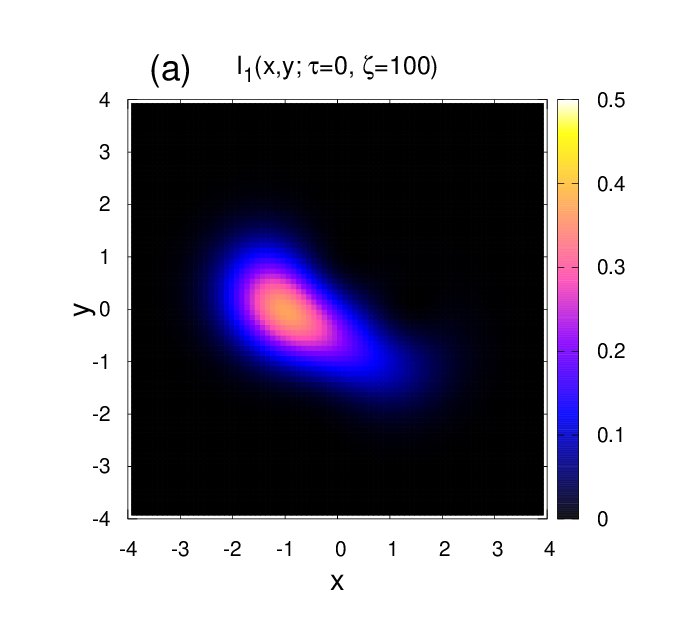, width=42mm}
\epsfig{file=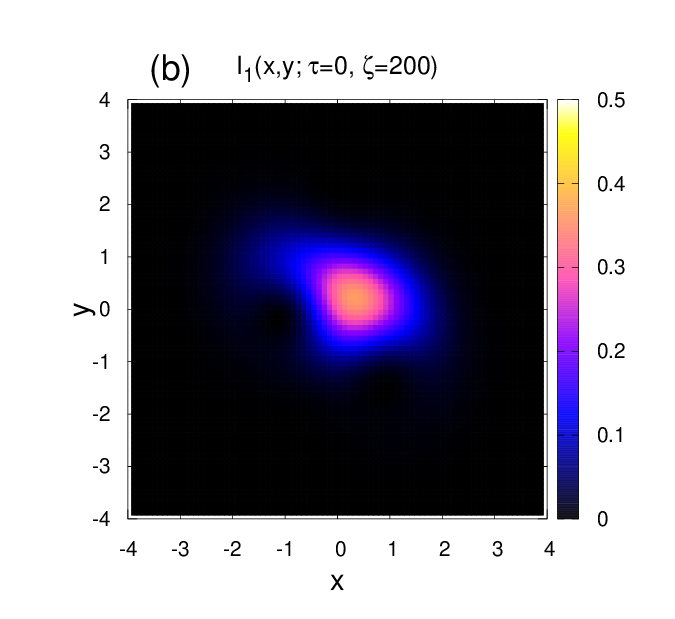, width=42mm}\\
\epsfig{file=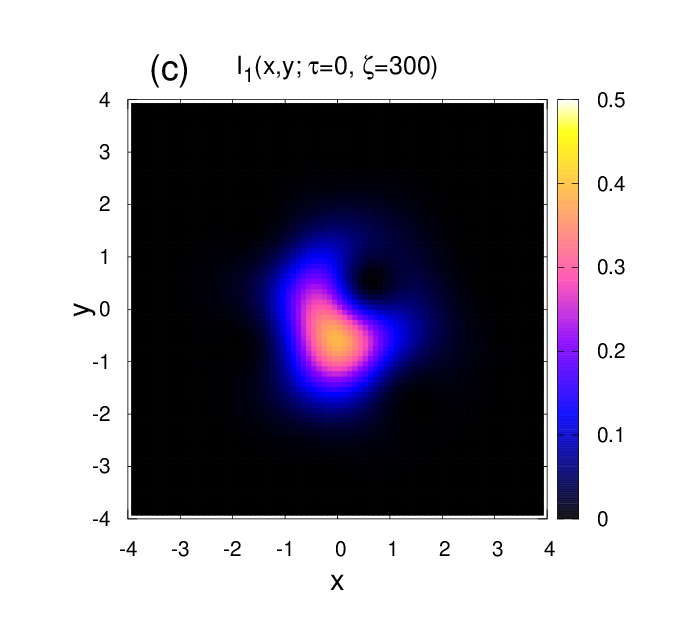, width=42mm}
\epsfig{file=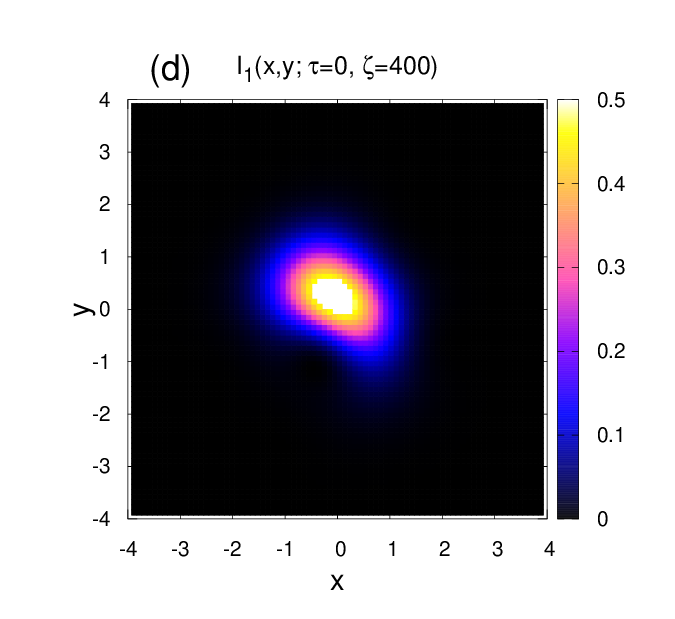, width=42mm}
\end{center}
\caption{Solitons ``merged'' after the tangential collision at 
$R=1.2$ and $v=0.1$. The $(x, y)$ map of the intensity $I_1(x,y)$ 
in the $\tau=0$ transverse plane at several propagation distances is shown. 
In this plane, $I_2(x,y)=I_1(-x,-y)$ due to the symmetry of the
initial conditions.
}
\label{tangent-collision-1} 
\end{figure}

\begin{figure}
\begin{center}
\epsfig{file=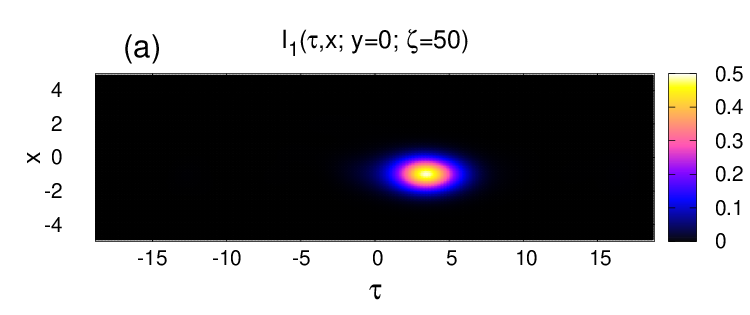, width=88mm}\\
\epsfig{file=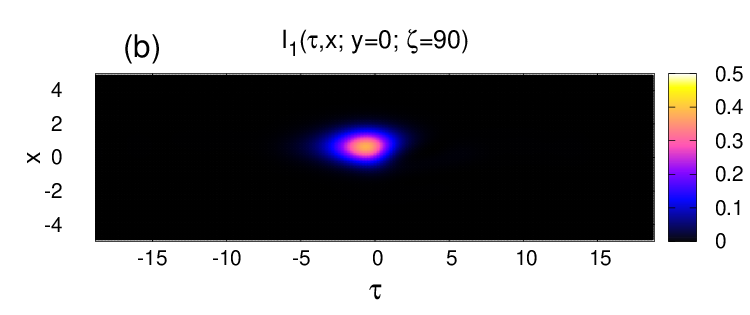, width=88mm}\\
\epsfig{file=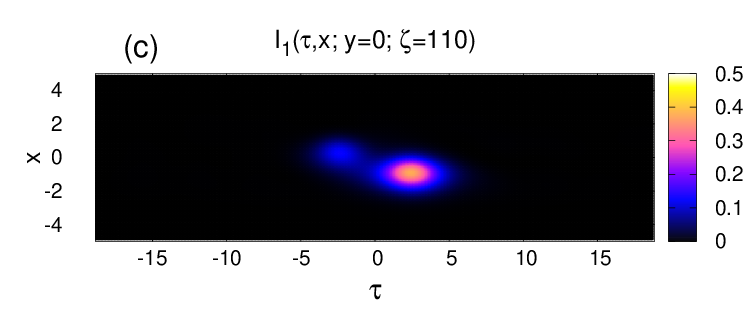, width=88mm}\\
\epsfig{file=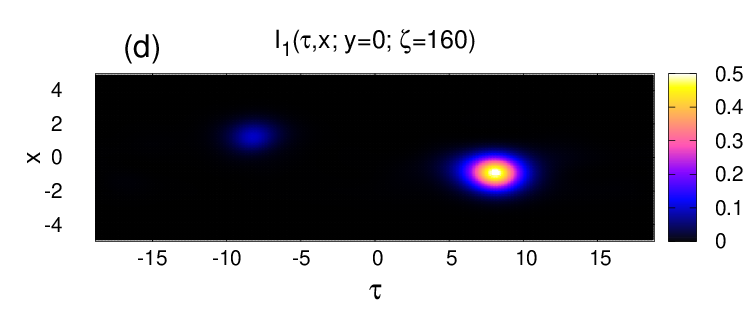, width=88mm}
\end{center}
\caption{Tangential collision at $R=1.0$ and $v=0.1$
resulting in the formation of two binary solitons.
The $(\tau,x)$ map of the intensity $I_1(\tau,x)$  in the $y = 0$
plane at several propagation distances is shown. In this plane, 
$I_2(\tau,x)=I_1(-\tau,-x)$ due to the symmetry of the initial 
conditions. Light bullets before and after the collision move 
along helical lines; for this reason, $\zeta$ coordinates at which
the centers of solitons are close to the intersection of the $y = 0$
plane have been chosen.
}
\label{tangent-collision-2} 
\end{figure}

However, significant differences were revealed in
other processes such as in collisions of two solitons (we
recall that the center of each of them moves on a classical 
trajectory so that they can closely approach in a
certain interval of the variable $\zeta$). Collisions of solitons
with opposite circular polarizations are significantly
different than collisions of identically polarized solitons. 
In particular, if the left and right solitons with
approximately equal numbers $N_1$ and $N_2$ approach
each other, the first soliton ``sees'' the second soliton
as an unfilled deep potential well for it. Naturally, the
first ``fluid'' tends to ``flow'' to this second well. 
Entering the second well, the first soliton transfers a part of
its momentum to it. Correspondingly, the second fluid
flows to the first well and both wells change their
shapes. The final result depends on the relative velocity 
of longitudinal approach $v$  and on the ``impact
parameter,'' i.e., the radius $R$ in terms of strictly helical
orbits of solitons in the axisymmetric potential. Thus,
even the minimum ``set'' includes four ``input''
parameters $N_1$, $N_2$, $v$ and  $R$. To ``trace'' the 
four-dimensional parametric region in sufficient detail, a
significant computer time is required. Only collisions
with $N_1=N_2=5.7$ (nonlinearity is not already weak,
but the soliton is still stable; the initial values
$a^2=b^2= 0.9$ and $c^2= 5.0$ are close to equilibrium)
have been already simulated more or less systematically.
Three values 0.1, 0.2, and 0.3 were taken for the
parameter $v$ and the radius $R$ was varied from 0.0 to 1.4
with a step of 0.2. Some characteristic results are presented 
in Figs. 2--4. Figure 2 shows the successive
stages of the head-on collision of two solitons with
opposite circular polarizations resulting in the formation 
of two binary light bullets. Figure 3 presents the
case $R = 1.2$ and $v = 0.1$, where tangential collision led
to the merging of two solitons into a more complex
rotating nonaxisymmetric structure. Figure 4 (where
$R = 1.0$ and $v = 0.1$) demonstrates the collision with
almost total reflection, where each soliton changed
the direction of its motion to opposite, acquiring only
a small fraction of the other component. This situation
differs from that presented in Fig.2, where each soliton 
predominantly continues its motion in the unchanged direction. 
Some collisions resulted in solitons with approximately equal
fractions of both components (e.g., at $R = 0.4$ and $v = 0.1$; 
this case is not shown).

Wave collapse occurred quite often, particularly
after secondary collisions near the edge of the computational 
domain (e.g., at $R = 0.2$ and $v = 0.1$; this case
is not shown). Several numerical experiments with
``heavy'' solitons at $N_1 = N_2 = 8.0$ were also carried
out. Such solitons have a low stability margin; therefore, 
collisions led to collapse in almost all cases except for quite
large radii $R\gtrsim 1.4$. This phenomenon is not discussed here.

The approach velocity was low in all reported cases,
and the processes of collision were rather long-term
and strongly inelastic. The collision of two fast light
bullets will be quasi-elastic.

Collisions of solitons with the identical circular
polarizations were also simulated for comparison. In
each of these cases, qualitative differences from the
examples reported above were observed, and the
development of events strongly depended on the initial
phase difference between single-component solitons.
For example, the head-on collision such as that presented 
in Fig.2 resulted in the immediate collapse at
zero phase difference, in the repulsion of solitons
before close approach at a phase difference of $\pi$, and in
the flow of the ``light fluid'' from one soliton to the
other at a phase difference of $\pi/2$ (these results are not
shown). The interaction of solitons with opposite circular
polarizations is independent of the phase.

\subsection*{Conclusions}

To summarize, the interaction between two polarizations 
of light has been taken into account for the first time when 
theoretically considering three-dimensional solitons in
graded-index multimode waveguides with the Kerr nonlinearity. 
The possibility of two-component light bullets has been shown. 
The numerical simulation of collisions of light bullets with
opposite circular polarizations has demonstrated
interesting diverse nonlinear dynamics significantly
different from the dynamics of collisions of identically
polarized solitons. The possible scenarios of the development
of events at the scattering of solitons are rich
and complex because of a larger number of the internal
degrees of freedom that are excited in the two-component 
system and interact with the translational motion
(similar to the one-dimensional case [24--26]). Only
several first steps have been already made in the study
of such binary structures. This field is certainly very
promising, in particular, in experiments because 
single-component solitons in the considered systems
have been already implemented in laboratories [5].

\subsection*{Funding}

This work was supported by the Ministry of Science and
Higher Education of the Russian Federation (state 
assignment no. FFWR-2024-0013).

\subsection*{Conflict of interests}

The author declares that he has no conflicts of interest.

\subsection*{Open access}

This article is licensed under a Creative Commons Attribution 
4.0 International License, which permits use, sharing,
adaptation, distribution and reproduction in any medium or
format, as long as you give appropriate credit to the original
author(s) and the source, provide a link to the Creative Commons 
license, and indicate if changes were made. The images
or other third party material in this article are included in the
article’s Creative Commons license, unless indicated otherwise 
in a credit line to the material. If material is not included
in the article’s Creative Commons license and your intended
use is not permitted by statutory regulation or exceeds the
permitted use, you will need to obtain permission directly
from the copyright holder. To view a copy of this license, visit
\url{http://creativecommons.org/licenses/by/4.0/}

\end{document}